\def \yskip{\penalty-50\vskip3pt plus 3pt minus 2pt}
\def \reference{\par \yskip \noindent \hangindent .4in \hangafter 1}
\def \abc#1#2#3#4 {\reference#1, {\sl#2}, {\bf#3}, #4}
\def \blank {\lower 5pt\hbox to 0.75in{\hrulefill}}

\def \cm{~\rm{cm}}
\def \s{~\rm{s}}
\def \km{~\rm{km}}

\def \g{~\rm{g}}
\def \AU{~\rm{AU}}

\def \yr{~\rm{yr}}

\def \lae{\mathrel{<\kern-1.0em\lower0.9ex\hbox{$\sim$}}}

\def \gae{\mathrel{>\kern-1.0em\lower0.9ex\hbox{$\sim$}}}

\documentstyle{article}

\begin{document}
\small


\title{ 
Axisymmetrical Structures of Planetary Nebulae and  
SN 1987A}

\author{
Noam Soker\\
Department of Physics, University of Haifa at Oranim\\
Oranim, Tivon 36006, ISRAEL \\
soker@physics.technion.ac.il }
\date{}
\maketitle

\begin{abstract}
 I summarize some recent models and ideas for the formation of 
axisymmetrical structures of planetary nebulae and the three rings of 
SN 1987A, as follows. 
 (a) I review the general role of binary companions, including brown dwarfs 
and planets. 
  (b) I propose a mechanism for axisymmetrical mass loss on the AGB 
that may account for the axially symmetric structures of elliptical 
planetary nebulae and that operates for slowly rotating AGB stars, 
$10^{-4} \Omega _{\rm Kep} \lae \Omega \lae 10^{-2} \Omega_{\rm Kep}$,
where $\Omega_{\rm Kep}$ is the equatorial Keplerian angular velocity.
(c) I propose a model for the formation of the two outer rings of SN 1987A,
which is based on the numerical simulation of Soker (1989),
and discuss a mechanism for their displacement from the exploding star. 
\end{abstract}

\noindent
{In the proceedings of the conference 
{\it Physical Processes in Astrophysical Fluids} (January 1998).
Will be published as a special volume of Physics Reports.}

\section{Introduction}

Scanning through recent images of SN 1987A (e.g. Burrows {\it et al.} 1995)
and through catalogs of planetary nebulae (PNs; e.g., Acker {\it et al.} 1992; 
Schwarz, Corradi, \& Melnick 1992; Manchado {\it et al.} 1996) 
we find that the circumstellar media of many stars at their final
nuclear burning phase have axisymmetrical, rather than spherical,
structures. 
 Axisymmetrical PNs which have two lobes with a morphological ``waist'' 
between them are termed ``bipolar PNs'' (also ``butterfly'' or ``bilobal''),  
while PNs which have a more elliptical than bilobal structure 
are termed elliptical PNs (Schwarz, Corradi, \& Stanghellini 1993). 
 The axisymmetrical structures of most PNs led to a 
debate on whether elliptical PNs can be formed through 
single-stellar evolution, or whether a binary companion is necessary 
(e.g., Fabian \& Hansen 1979;  Livio 1982, 1998; 
Livio, Salzman, \& Shaviv 1979; Webbink 1979; Morris 1981; 
Zuckerman \& Gatley 1988; Pascoli 1992; Iben \& Livio 1993; 
Soker 1997, 1998a; Balick {\it et al.} 1994; Pottasch 1995; 
Pollacco \& Bell 1997; Corradi {\it et al.} 1996;
Kastner {\it et al.} 1996). 
 In the last decade this debate was extended to the formation of the
nonspherical explosion and three rings of SN 1987A. 
In many PNs, as well as in the three rings of SN 1987A, 
there are displacements of the nebulae from the central stars,
which hint at the interaction of the progenitors with wide binary 
companions, with close binaries having eccentric orbits, or with the ISM
($\S 5$).
 
 In a recent paper (Soker 1997) I suggest that four main 
evolutionary routes determine the degree of asymmetry of the 
axially symmetric structures of PNs. 
 I then classify 458 PNs according to the 
process which caused their progenitors to blow axisymmetrical winds. 
 The classification is based primarily on the morphologies of the different 
PNs, assuming that binary companions, stellar or substellar,
are necessary for axisymmetrical mass loss on the AGB.
 The four evolutionary classes, 
according to the binary-model hypothesis, are: 
\newline
(a) Progenitors of planetary nebula which did not
interact with any companion, and therefore they rotate extremely slowly
when reaching the AGB. 
These amount to $\sim 10 \%$ of all planetary nebulae. 
\newline
(b) Progenitors which interact with stellar companions which 
avoided a common envelope, $11 ^{+ 2}_{-3} \%$ of all nebulae.  
These form bipolar PNs, as is the case in symbiotic nebulae
(Morris 1990; Schwarz \& Corradi 1992; Soker 1998a). 
\newline
(c) Progenitors which interact with stellar companions via a common envelope
phase, $23^{+11}_{-5} \%$ of all nebulae.  
These form extremely asymmetrical structures, i.e., tori, elongated
elliptical PNs, and in some cases bipolar PNs. 
\newline
(d) Progenitors which interact with {\it substellar} (i.e., planets and brown 
dwarfs) companions via a common envelope phase, $56^{+5}_{-8} \%$ 
of all nebulae.  
 These form elliptical PNs with relatively small deviation from sphericity.
\newline
 These numbers are compatible with other studies 
(e.g., Yungelson, Tutukov, \& Livio 1993; 
Han, Podsiadlowski, \& Eggleton 1995). 

 In $\S 2$ I discuss the problem of angular momentum of AGB stars, 
which suggests that to account for the $\sim 60 \%$ elliptical 
PNs either there are many planetary systems ($\S 3$; Soker 1996; 1997) 
or there is a mechanism to induce axisymmetrical mass loss from
very slowly rotating AGB stars.
 Such a model for singly evolved very slowly rotating AGB stars 
is the mechanism of mode-switch to nonradial oscillations,  
proposed by Soker \& Harpaz (1992). 
 In $\S 4$ I propose yet another model (Soker 1998c) which
may operate in singly evolved AGB stars.
  This model is based on both magnetic activity and radiation pressure on 
dust.
 In $\S 5$ I propose a model for the two outer rings of SN 1987A.

\section{Angular Momentum Considerations}

 If most of the mass loss occurs on the AGB, the ratio of envelope 
angular velocity on the upper AGB to the Keplerian 
(critical) angular velocity for a single star evolution is given by 
(Soker 1998c)
\begin{eqnarray}
\left( {{\Omega}\over{\Omega_{\rm Kep}}} \right)_{\rm Single-AGB}
\simeq 
10^{-4}     
\left( {{\Omega}\over{0.1 \Omega_{\rm Kep}}} \right)_{\rm MS}
\left( {{R_{\rm MS}}}\over{0.01R_{\rm {AGB}}} \right)^{1/2} 
\left({{M_{\rm {env}}} \over {0.1 M_{\rm {env0}}}} \right)^ {2},
\end{eqnarray}
where the subscript MS means that the quantity is taken
at the end of the main sequence, and $M_{\rm {env0}}$ is the
envelope mass at the beginning of the AGB. 
 A faster rotation on the AGB can be attained if a binary companion, stellar 
 or substellar, spins-up the envelope.
  The orbital separation of a low mass secondary when tidal interaction 
becomes significant is $a \sim 5 R_\ast$ (Soker 1998c), where $R_\ast$
is the stellar radius.
 If the secondary deposits all its orbital angular momentum to the
envelope of mass $M_{\rm env}$, the ratio of envelope angular velocity 
$\Omega$ to the surface Keplerian angular velocity $\Omega_{\rm Kep}$ 
is given by 
\begin{eqnarray}
{{\Omega}\over{\Omega_{\rm Kep}}} \simeq 
0.1   
\left( {{M_2}\over{0.01M_{\rm env}}} \right) 
\left({{a} \over {5R_\ast}} \right)^ {1/2},
\end{eqnarray}
assuming that the entire envelope rotates uniformly and has a density 
profile of $\rho_{\rm env} \propto r^{-2}$. 
From the last two equations it is clear that for any model that requires 
AGB angular velocity of more than $\sim 10^{-3} \Omega_{\rm Kep}$, 
a substantial spin-up is required. 

 The dust-based model for axially symmetric mass loss proposed by 
Dorfi \& H\"ofner (1996) requires an AGB star of radius $R = 500 R_\odot$ 
to rotate at $\gae 10 \%$ of the Keplerian angular velocity. 
 We conclude that in order to spin-up the envelope as required by 
Dorfi \& H\"ofner the secondary mass should be 
$M_2 > 0.01 M_{\rm {env}}$. 
 However, as Harpaz \& Soker (1994) show, the envelope's specific angular
momentum of an AGB star decreases with mass loss as 
$L_{\rm env}/M_{\rm env} \propto M_{\rm env}^{2}$. 
 Therefore, to supply the angular momentum for a longer time, 
the companion mass should be much larger than $0.01 M_\odot$, i.e.,
a brown dwarf or a low main sequence star. 
 
 Direct magnetic effects, through magnetic tension and/or pressure,
have been suggested to determine the mass loss geometry from AGB stars 
(Pascoli 1997), or to influence the circumstellar
structure during the PN phase (Chevalier \& Luo 1994; Garcia-Segura 1997).
 These models {\it must} also incorporate a binary companion to substantially 
spin-up the envelope (Soker 1998c). 
 The model of Chevalier \& Luo (1994) is based on the tension of the toroidal component of the 
magnetic field in the wind: the wind in the transition from the AGB to the
PN phase or the fast wind during the PN phase. 
 Close to the star the magnetic pressure and tension are negligible compared
with the ram pressure and thermal pressure of the wind. 
 As the wind hits the outer PN shell, which is the remnant of the slow wind,
it goes through a shock, slows down and the toroidal component 
of the magnetic field increases substantially. 
 This may result in the magnetic tension and pressure becoming the 
dominant forces near the equatorial plane.   
 The efficiency of this model is determined by a parameter given by
(Chevalier \& Luo 1994) 
\begin{equation}
\sigma
=
\left( {{B_s^2r_s^2}\over{\dot M_w v_w}} \right)
\left( {{v_{\rm {rot}}}\over{v_w}} \right)^2 
= {{\dot E_B}\over{\dot E_k}} 
\left({{v_{\rm {rot}}}\over{v_w}} \right)^2, 
\end{equation}
where $B_s$ is magnetic field intensity on the stellar surface,
$r_s$ the stellar radius,  $\dot M_w$ the mass loss rate into the wind, 
$v_w$ the terminal wind velocity, and $v_{\rm {rot}}$ the equatorial 
rotational velocity on the stellar surface.
  In obtaining the second equality the expressions for the magnetic energy 
luminosity  $\dot E_B=4 \pi r_s^2 v_w B_s^2 / 8 \pi$ and for the kinetic 
energy luminosity $\dot E_k = \dot M_w v_w^2/2$ were used. 
 For the model to be effective it is required that $\sigma \gae 10^{-4}$,
but a typical value of $\sigma \simeq 0.01$ is used by 
Garcia-Segura (1997). 
 Since the magnetic field is weak near the star $\dot E_B \lae \dot E_k$,
the model requires, by equation (3),  $v_{\rm {rot}} \gae 0.01 v_w$.  

 From this discussion it turns out that to account
for the $\sim 60 \%$ elliptical PNs we have to adopt one of the following.
 Either there are many planetary systems (Soker 1996; 1997; next section) 
or there is a mechanism to induce axisymmetrical mass loss from
very slowly rotating AGB stars
(Soker \& Harpaz 1992; Soker 1998c; $\S 4$ below). 

\section{Planets}

  As mentioned in $\S 1$, if spin-up is required for axisymmetrical mass 
loss, then $\sim 1/2$ of all PN progenitors are influenced by
plant or brown dwarf companions. 
 Based on this conclusion I argued (Soker 1996) that substellar objects 
(brown dwarfs or gas-giant planets) are commonly present within 
several$\times$AU around main sequence stars.  
For a substellar object to have a high probability of being present within 
this orbital radius, on average several substellar objects must be 
present around most main sequence stars of masses $\lae 5 M_\odot$. 
 This led me to suggest that the presence of four gas-giant 
planets in the solar system is typical.
 
 As a star evolves along the RGB or the AGB its radius increases.
 Any close planet will eventually interact tidally with the star.
 Since a substellar companion cannot bring the envelope to corotation,
it will spiral-in to form a common envelope. 
 This happens when the tidal interaction time is shorter than 
$\tau_{\rm ev}$, the time spent by the star on the RGB or AGB. 
 This condition gives the maximal tidal interaction orbital separation in 
Zahn's (1989) equilibrium tide model, for $M_2 \ll M_1$ 
and neglecting weak dependences on the luminosity and on the radius (Soker 1996), 
\begin{eqnarray}
a_{\rm max} \simeq 4 R_\ast  
\left( {{\tau_{\rm ev}}\over{6\times10^5 \yr}} \right)^{1/8} 
\left( {{M_{\rm {env}}} \over {0.5M_1}} \right)^ {1/8} 
\left( {{M_{\rm {env}}} \over {0.5M_\odot}} \right)^ {-1/24} 
\left( {{M_2} \over {0.01M_1}} \right)^ {1/8}, 
\end{eqnarray}
where $M_{\rm {env}}$ is the envelope mass. 
Stellar quantities are taken at the RGB tip and 
AGB tip, for RGB and AGB stars, respectively. 
 Other effects of planets on AGB stars are summarized in 
Soker (1997). 
 Planets with orbital separation of $\gae 5$ times the
maximum radius of a star on the AGB will not enter the envelope.
 They will survive to the PN phase, and if they are closer than 
$\sim 20 \AU$ they will be strongly ionized by the central star,
and may reveal themselves as compact ionized high density regions 
(Soker \& Dgani 1998). 

 Low mass main sequence stars $M \lae 2 M_\odot$ swell to large radii
already on the RGB.  
 Therefore they are likely to interact with their close planets
on the RGB. This will influence their subsequent location on the
horizontal branch. 
 Planets, therefore, may cause some anomalies on the horizontal branch 
of globular clusters, and may be related to the second parameter
of the horizontal branch (Soker 1998b). 
 
\section{AGB Stellar Spots and Dust Formation}

 If in the next few years the results of the intensive 
planet search projects (Marcy \& Butler 1998)
are that only $\ll 50 \%$ of all stars have planets, then  
model will be needed of efficient axisymmetrical mass loss for singly 
evolved stars. 
 In addition, the mechanism should account for the increase in the
degree of asymmetry toward the termination of the AGB evolution 
(Soker 1997), which is observed in many PNs.  
Such a mechanism, based on mode-switch to nonradial oscillations,  
was proposed by Soker \& Harpaz (1992). 
 Recently I proposed a different mechanism (Soker 1998c), which  
may operate for slowly rotating AGB stars, having angular velocity 
in the range of
$10^{-4} \Omega _{\rm Kep} \lae \Omega \lae 10^{-2} \Omega_{\rm Kep}$,
where $\Omega_{\rm Kep}$ is the equatorial Keplerian angular velocity.
 Such angular velocities could be gained from a planet companion of
mass $\gae 0.1 M_{\rm Jupiter}$, which deposits its orbital angular
momentum to the envelope during the AGB phase or even much earlier
during the RGB phase, or even from single stars which are fast 
rotators on the main sequence.   
 The proposed model incorporates both dynamo magnetic activity and radiation
pressure on dust. 
 The magnetic activity results in the formation of cool spots, above which
dust forms much easily.  
 The enhanced magnetic activity toward the equator results in a higher
dust formation rate there, hence higher mass loss rate. 
 As the star ascends the AGB, both the mass loss rate and magnetic activity
increase rapidly. 
 The model is built to explain elliptical PNs, but not the more
extremely asymmetrical bipolar PNs, which are thought to be formed from 
stellar binary systems. 

\section{The Rings Around SN 1987A}

 I start by presenting a 2D numerical simulation I performed 
(Soker 1989) as a speculative effect for shaping proto-PNs. 
 As noted already by Soker \& Livio (1989), models for the formation
of axisymmetrical PNs may be relevant to SN 1987A as well. 
 In that paper I assumed that in the transition from the AGB to the PN, 
the star has a short mass loss episode, a ``pulse'', due to an interaction 
with a binary companion.  
 In this pulse, mass loss occurs close to the equatorial plane, and at a 
velocity faster than that of the slow wind. 
 In that specific simulation I assumed that the pulse occurs 600 
years after the end of the slow wind, it lasts for 50 years, and it is 
concentrated within an angle of $\sim 10 ^\circ$ from the 
equatorial plane.
 The velocity of the material in the pulse is $200 \km \s^{-1}$,
and the total mass $3 \times 10^{-4} M_\odot$.
 This pulse runs into the slow wind, which has a density contrast of 
6 between the equator (high density) and polar directions, 
a mass loss rate of $10^{-5}  M_\odot \yr^{-1}$, and
a velocity of $10 \km \s^{-1}$.
 As the pulse hits the slow wind, a high pressure region is formed in 
a small region in the equatorial plane (in 3D it has the shape of a ring).
The fast release of energy resembles an explosion, and it creates a shell
expanding from the high pressure region.  
 Since there is no slow wind material inward, the shell has the shape
of a horseshoe as observed in the symmetry plane 
(the plane perpendicular to the equatorial plane). 
 The density contours and velocity field map in the symmetry plane
570 years after the pulse reaches the slow wind are presented in Figure 1.
 For SN 1987A different parameters should be used, so this figure
should be applied qualitatively rather than quantitatively. 
 The ``horseshoe'' shape is the projection of a 3D hollowed torus 
on the symmetry plane. 
It should be emphasized that the qualitative result, of forming a 
horseshoe-torus, will still hold for a much longer ``pulse'', and it 
does not depend on a density contrast between the equator and poles. 
 We note that the inner region of the ``horseshoe'' in Figure 1
is 
(a) denser than most of the other regions of the torus, 
(b) its velocity is the lowest, and
(c) it is extended in a radial direction more or less. 

 Assume now that the central star starts to blow a fast wind, which
hits the horseshoe-torus and accelerates it. 
 The inner region of the horseshoe-torus will have the lowest acceleration
since it is denser, and more over, it is elongated in a more or less
radial direction.
  Therefore, there will be a time in the evolution of this flow when
most of the original torus has already been expelled to large distances 
from the star by the fast wind, and hence it has low density. 
 The original inner regions of the torus, by contrast, will expand slower, 
be relatively close to the star, and be dense, much denser than the 
wind material around it.
 These regions now form two rings, one at each side of the equatorial plane. 
 
 Let me try to sketch a scenario for the formation of the three
rings around SN 1987A based on the discussion above. 
 One assumption that goes here and in the explanation for the displacement 
of the rings from the exploding star is that of a binary companion in an
initial eccentric orbit.
 There are other reasons to support a binary companion. 
 First, Chevalier \& Soker (1989) show that a deformed envelope due to fast 
rotation can explain the asymmetrical explosion of SN 1987A, which is inferred 
from polarization data. 
 The direction of polarization is in the direction of the symmetry axis 
of the ring (or perpendicular to it, depending on the time and emission
lines). 
 The angular velocity required can be gained from a companion of mass
$\gae 0.5 M_\odot$. 
 Second, the merger of a $\sim 5 M_\odot$ secondary with the progenitor's 
envelope makes the envelope shrink, hence the transition of
the progenitor to a blue star (Podsiadlowski, Joss, \& Rappaport 1990). 

 There are two possible tracks to the proposed evolutionary sequence, 
 marked A and B here.
  In this preliminary study, I cannot prefer one or the other.
 The different stages of the proposed scenario are depicted in Figure 2. 
\newline  
  {\bf phase 1:} Slow wind from the red supergiant progenitor.
 It may have a higher mass loss rate in the equatorial plane due to
rotation, or a tidal interaction with the companion.
(A) This wind is displaced relative to the central star and perpendicular
to the symmetry axis, due to an interaction with the companion
which is in an eccentric orbit (Soker, Harpaz, \& Rappaport 1998). 
 This explains the displacement of the rings of SN 1987A from the central
 star. 
(B) The companion still has a weak influence. 
\newline
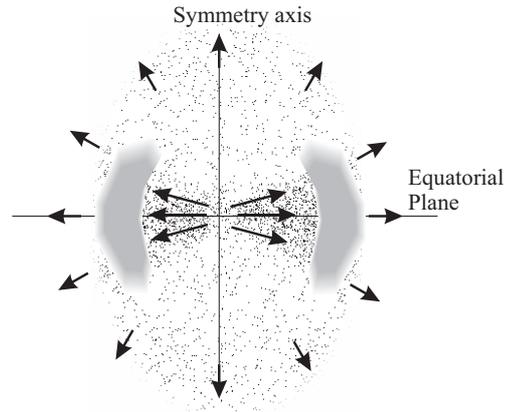
{\bf phase 2:} The binary system blows a faster wind concentrated in the 
equatorial plane.  A horseshoe-torus is formed. 
(A) The fast winds occurs as the companion approaches the primary 
(the progenitor of 1987A) and enters its envelope. 
(B) The fast wind results from a strong interaction with the companion, 
and it is displaced due to the eccentric orbit (Soker {\it et al.} 1998). 
The eccentricity of the orbital motion of the companion results in the
displacement of this faster wind perpendicular to the symmetry axis, and 
later leads to the displacement of the three rings.
\newline
{\bf phase 3:} A slow dense wind is blown in the equatorial plane. 
 The inner edge of this wind forms the inner ring of SN 1987A.
(A) The slow wind concentration in the equatorial plane is due to the
fast rotation of the envelope, after the companion enters the envelope
but before the primary's shrinkage to a blue giant.
 This lasts several thousand years (Podsiadlowski {\it et al.} 1990). 
(B) Most of this phase occurs while the secondary is still outside the 
envelope, and it ends several thousand years after the secondary enters the
primary's envelope. 
\newline
{\bf phase 4:} Several thousand years after the secondary, 
both for tracks A and B, enters the envelope, the primary shrinks to a blue
giant (Podsiadlowski {\it et al.} 1990), and blows a fast wind.  
 This wind pushes most of the previous dense wind to large distances, beside
the very dense regions which are also elongated in radial directions: 
the inner regions of the horseshoe-torus and the dense disk in the 
equatorial plane. 
\newline 
{\bf phase 5:} Just before explosion the system contains three rings:
 two outer rings, one at each side of the equatorial plane, which 
are the remnant of the inner regions of the horseshoe-torus,
and a inner dense ring in the equatorial plane. 

Several comments should be made here.
{\bf (a)} The proposed scenario accounts for: (i) slow motion of the outer 
rings; (ii) the high density contrast of the outer rings to their 
surroundings; (iii) the presence of only two outer rings;
(iv) the displacement of the three rings relative to the central star. 
{\bf (b)} Phase 2 is the most problematic in the proposed scenario.
 That is, how come the binary system blows an equatorial wind of velocity 
$\sim 100 \km \s^{-1}$ before the primary shrinks? 
 The support for such a wind, which is not found in PNs, is that both the
 progenitor and the companion are more massive than in PNs. 
 A companion of $\sim 5 M_\odot$ has a wind much stronger and faster than 
the expected companions of bipolar PNs.  
 It is possible that the wind of the companion is forced to the equatorial
plane, and is responsible for such a faster wind. 
Observationally, there is a collimated flow in the equatorial plane
(but only in one specific direction) of Eta Carinae, despite the nice 
two lobes and equatorial waist of Eta Carinae. 
 The equatorial wind, and the other phases, are currently being studied
by us (Soker \& Rappaport 1998). 
{\bf (c)} All three rings are displaced  in the same direction relative
to the exploding star, though the outer two rings are more displaced. 
 Three process can cause displacement of circumstellar 
nebulae (Soker 1997): ($i$) interaction with the ISM: 
this is not likely here because it cannot influence the inner ring
and it will deform the outer rings; 
($ii$) interaction with a wide companion, 
having an orbital period of several$\times 10^4 \yr$; 
($iii$) an eccentric close companion (Soker {\it et al.} 1998), 
as we proposed is the case for SN 1987A. 
{\bf (d)} The proposed scenario predicts that there is matter extended outward
to the two outer rings in the radial direction, and that there is matter
from the broken shell between the two rings, but at much greater
distances from the exploding star. 
{\bf (e)} There are other models for the formation of the rings. 
Burrows {\it et al.} (1995) summarize several models, 
and find problems in all of them.
 Meyer (1997) proposes that the rings are formed from ionization that 
induces hydrodynamic motions. 
 I find two problems in this model.
 First the density contrast seems to be too low in it, and second,
the inner ring would prevent the ionization of the equatorial region between 
the two outer rings. 
 The last effect will reduce the efficiency of the model, reducing
further the density contrast. 

{\bf ACKNOWLEDGMENTS:} 
I thank Saul Rappaport for several helpful comments.
 This research was in part supported by a grant from the 
Israel Science Foundation. 

\section*{References}

\reference{}Acker, A., Ochsenbein, F., Stenholm, B., Tylenda, R., Marcout, J.,  
\& Schohn, C. 1992, Strasbourg-ESO Catalogue of Galactic Planetary Nebulae
(ESO).  

\reference{}Balick, B., Perinotto, M., Maccioni, A., Terzian, Y., \& 
Hajian, A. 1994, ApJ, 424, 800.

\reference{}Burrows, C. J. {\it et al.} 1995, ApJ, 452, 680. 

\reference{}Chevalier, R. A., \& Luo, D. 1994, ApJ, 421, 225. 

\reference{}Chevalier, R. A., \& Soker, N. 1989, ApJ, 341, 867.

\reference{}Corradi, R. L. M., Manso, R., Mampaso, A.,
\& Schwarz, H. E. 1996, A\&A, 313, 913.

\reference{}Dorfi, E. A., \& H\"ofner, S. 1996, A\&A, 313, 605.

\reference{}Fabian, A. C., \& Hansen, C. J. 1979, MNRAS, 187, 283. 

\reference{}Garcia-Segura, G. 1997, ApJ, 489, L189. 

\reference{}Han, Z., Podsiadlowski, P., \& Eggleton P. P. 1995,
MNRAS, 272, 800. 

\reference{}Harpaz, A., \& Soker, N. 1994, MNRAS, 270, 734. 

\reference{}Iben, I. Jr., \& Livio, M. 1993, PASP, 105, 1373. 

\reference{}Kastner, J. H., Weintraub, D. A., Gatley, I., Merrill, K. M., 
\& Probst, R. G. 1996, ApJ, 462, 777.

\reference{}Livio, M. 1982, A\&A, { 105}, 37.

\reference{}Livio, M. 1998,
in Planetary Nebulae, eds. H. Lamers \& H. J. Habing, in press. 

\reference{}Livio, M., Salzman, J., \& Shaviv, G. 1979, MNRAS, 188, 1.

\reference{}Manchado, A., Guerrero, M., Stanghellini, L., 
\& Serra-Ricart, M. 1996, The IAC Morphological Catalog of Northern 
Galactic Planetary Nebulae. 

\reference{}Marcy, G., \& Butler, R. P. 1998, Sky \& Telescope, 
March p. 30.

\reference{}Meyer, F. 1997, MNRAS, 285, L11. 

\reference{}Morris, M. 1981, ApJ, 249, 572. 

\reference{}Morris, M. 1990, 
in { From Miras to PNs: Which
Path for Stellar Evolution?} eds. M.O. Mennessier \& A. Omont
(Paris: Ed. Frontieres), 520. 

\reference{}Pascoli, G. 1992, PASP, 104, 350. 

\reference{}Pascoli, G. 1997, ApJ, 489, 946. 

\reference{}Podsiadlowski, Ph., Joss, P. C., \& Rappaport, S. 1990, 
A\&A Lett., 227, L9. 

\reference{}Pollacco, D., \& Bell, S. A. 1997, 
MNRAS, 284, 32. 

\reference{}Pottasch, S. R. 1995,   
Ann. of the Israel Physical Society, Vol. 11: 
{ Asymmetrical Planetary Nebulae}, 
eds. A. Harpaz \& N. Soker (Haifa, Israel), p. 7. 

\reference{}Schwarz, H. E., \& Corradi, R. L. M. 1992, A\&A, 265, L37. 

\reference{}Schwarz, H. E., Corradi, R. L. M., \& Melnick, J. 1992, 
A\&AS, 96, 23. 

\reference{}Schwarz, H. E., Corradi, R. L. M, \& Stanghellini, L. 1993,
in Planetary Nebulae, IAU Symp. N155, eds. R. Weinberger and A. Acker
(Kluwer), p. 214. 

\reference{}Soker, N. 1989, ApJ, 340, 927. 

\reference{}Soker, N. 1996, ApJL, 460, L53. 

\reference{}Soker, N. 1997,  ApJSupp, 112, 487.  

\reference{}Soker, N. 1998a,  ApJ, 496, 833.  

\reference{}Soker, N. 1998b,  preprint.   

\reference{}Soker, N. 1998c,  preprint.   

\reference{}Soker, N., \& Dgani, R. 1998, preprint 

\reference{}Soker, N., \& Harpaz, A. 1992, PASP, 104, 923. 

\reference{}Soker, N., Harpaz, A., \& Rappaport, S. A. 1998, ApJ, in press. 

\reference{}Soker, N., \& Livio, M. 1989, ApJ, 339, 268. 

\reference{}Soker, N., \& Rappaport, S.  1998, in preparation.

\reference{}Webbink, R.F. 1979, { IAU Colloquium 46, Changing Trends
in Variable . . . }, ed. F.M. Bateson, J. Smak, \& I.H. Urch
(Hamilton, New Zealand), p. 102.

\reference{}Yungelson, L. R., Tutukov, A. V., \& Livio, M. 1993, 
ApJ, 418, 794. 

\reference{}Zahn, J-P. 1989, A\&A, 220, 112. 

\reference{}Zuckerman, B., \& Gatley, I. 1988, ApJ, 324, 501.  

\clearpage

$$
$$

$$
$$
\centerline {\it This figure is Figure 4 of Soker  (1989). }
\centerline {\it Contact Noam Soker
for the figure to be sent to you via mail or FAX.}

\noindent {\bf Figure 1:} The projection of the horseshoe-torus on the symmetry 
plane. The $Y$ axis is the symmetry axis and $X$ axis is in the equatorial
plane. The units on the axes are $10^{15} \cm$. 
 Each unit length (as measured on the axes) of the arrows corresponds to
a velocity of $12 \km \s^{-1}$, and the density levels, in units of
$10^{-21} \g \cm^{-3}$, are 4, 8, 16, 24, 32, 40, 48, 56. 

$$
$$

$$
$$

\centerline {\it The postscript of this figure is attached.} 
\centerline {\it Contact Noam Soker for this figure to be sent to you via 
mail or FAX.} 

\noindent {\bf Figure 2:} Schematic illustration of the proposed scenario
for the formation of the three rings of SN 1987A.

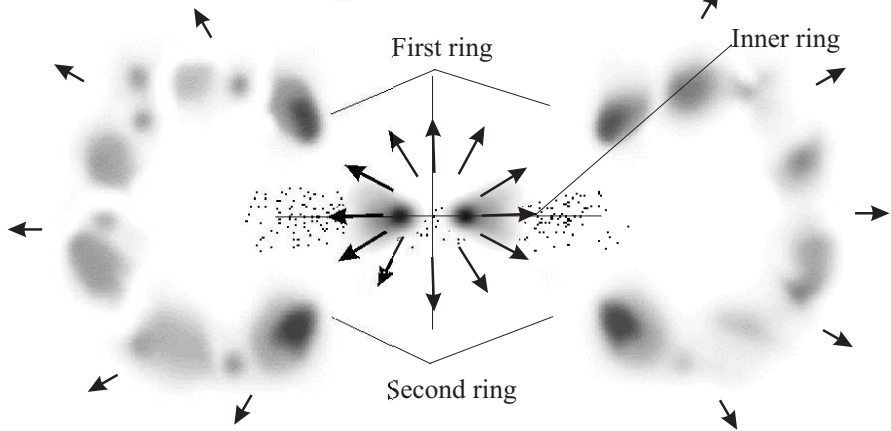
\end{document}